\documentclass[
    ,final            
  ]
  {aipproc}

\layoutstyle{6x9}

\begin{document}

\title{A Multicoloured View of 2S 0114+650}


\author{S.A. Farrell}{
  address={School of PEMS, UNSW@ADFA, Canberra ACT 2600, Australia}
}

\author{P.M. O'Neill}{
  address={Astrophysics Group, Imperial College, London SW7 2AZ, UK} }

\author{R.K. Sood}{
  address={School of PEMS, UNSW@ADFA, Canberra ACT 2600, Australia}
}

\author{S. Dieters}{
  address={School of Physics, University of Tasmania, Hobart TAS 7001, Australia}
}

\author{R.K. Manchanda}{
  address={Tata Institute of Fundamental Research, Mumbai 400005, India}
}

\begin{abstract}
We report the results of radio and X-ray observations of the high
mass X-ray binary 2S 0114+650, made with the Giant Meterwave Radio
Telescope and the $\textit{Rossi X-ray Timing Explorer}$
respectively. No emission was detected at radio wavelengths. The
neutral hydrogen column density was found to vary over the orbital
period, while no variability over the the super-orbital period was
observed. We discuss the causes of the observed relationships and
the implications for the underlying mechanisms.

\end{abstract}

\maketitle


\section{Introduction}

2S 0114+650 was discovered in X-rays in 1977, with the optical donor
star LSI +65$^{\circ}$ 010 being identified shortly after and
classified as a B1 Ia supergiant at a distance of 7.0 $\pm$ 3.6 kpc
\cite{Reig:1996}. This classification was recently verified through
near infrared observations at the Mt Abu observatory
\cite{Ashok:2006}. The binary orbital period of 11.6 d was
determined in X-rays using data from the All Sky Monitor (ASM)
aboard the \textit{Rossi X-ray Timing Explorer (RXTE)}
\cite{Corbet:1999}. An X-ray periodicity at $\sim$2.7 h was also
confirmed and attributed to emission from a highly magnetised
neutron star, one of the slowest known for an X-ray pulsar. The
discovery of a 30.7 d super-orbital period in ASM data was recently
reported \cite{Farrell:2006}. This period has been stable over
$\sim$8.5 yr making 2S 0114+650 the fourth X-ray binary exhibiting a
stable super-orbital modulation \cite{Sood:2006}. The underlying
mechanism behind this modulation is yet to be determined.

Here we present the results of complementary radio and X-ray
observations of 2S 0114+650 made with the Giant Meterwave Radio
Telescope (GMRT) and the $\textit{RXTE}$ satellite.

\section{Radio Observations}

Between 2005 January 3 -- 2005 February 20 we observed 2S 0114+650
with the GMRT near Pune, India. A total of 10 observations were
performed at 240, 321, 618, 1276 and 1386 MHz each with 16 MHz
bandwidth. The aim of these observations was to determine whether
there is any detectable radio emission from this source and if so
whether it is modulated over the spin, orbital or super-orbital
periods. The data were reduced using the 31DEC04 version of AIPS and
radio maps were produced. Due to the absence of sufficiently bright
sources in the field of view self calibration could not be performed
on the 1276 and 1386 MHZ images. The radio sources present in the
field all coincide with known radio sources in the NVSS catalogue
\cite{Condon:1998}. No radio emission was detected in any of the
observed bands within the known vicinity of 2S 0114+650. The flux
was corrected for the shape of the primary beam and the RMS noise
measured for the area containing 2S 0114+650. The 3$\sigma$ flux
upper limits were thus determined. Significant interference was
encountered in the 240 MHz band rendering channels 50 -- 128
unusable. Thus the 240 MHz limit is significantly higher than the
value theoretically obtainable. Table~\ref{tab1} lists the 3$\sigma$
flux upper limits for 2S 0114+650 for each of the frequency bands
observed.

\begin{table}[h!b]
\begin{tabular}{ccc}
\hline \tablehead{1}{c}{b}{Frequency\\(MHz)}
  & \tablehead{1}{c}{b}{Upper Limit\\(mJy)} & \tablehead{1}{c}{b}{Time on Source\\(hr)}\\
\hline
240 & 6.2 & 7.3\\
321 & 2.7 & 2\\
618 & 1.7 & 7.3\\
1276 & 0.9 & 1.4\\
1386 & 1.1 & 1.4\\
\hline
\end{tabular}
\caption{3$\sigma$ flux upper limits from the GMRT} \label{tab1}
\end{table}


\section{X-ray Observations}

\begin{figure}[h!]
  \includegraphics[height=0.35\textheight]{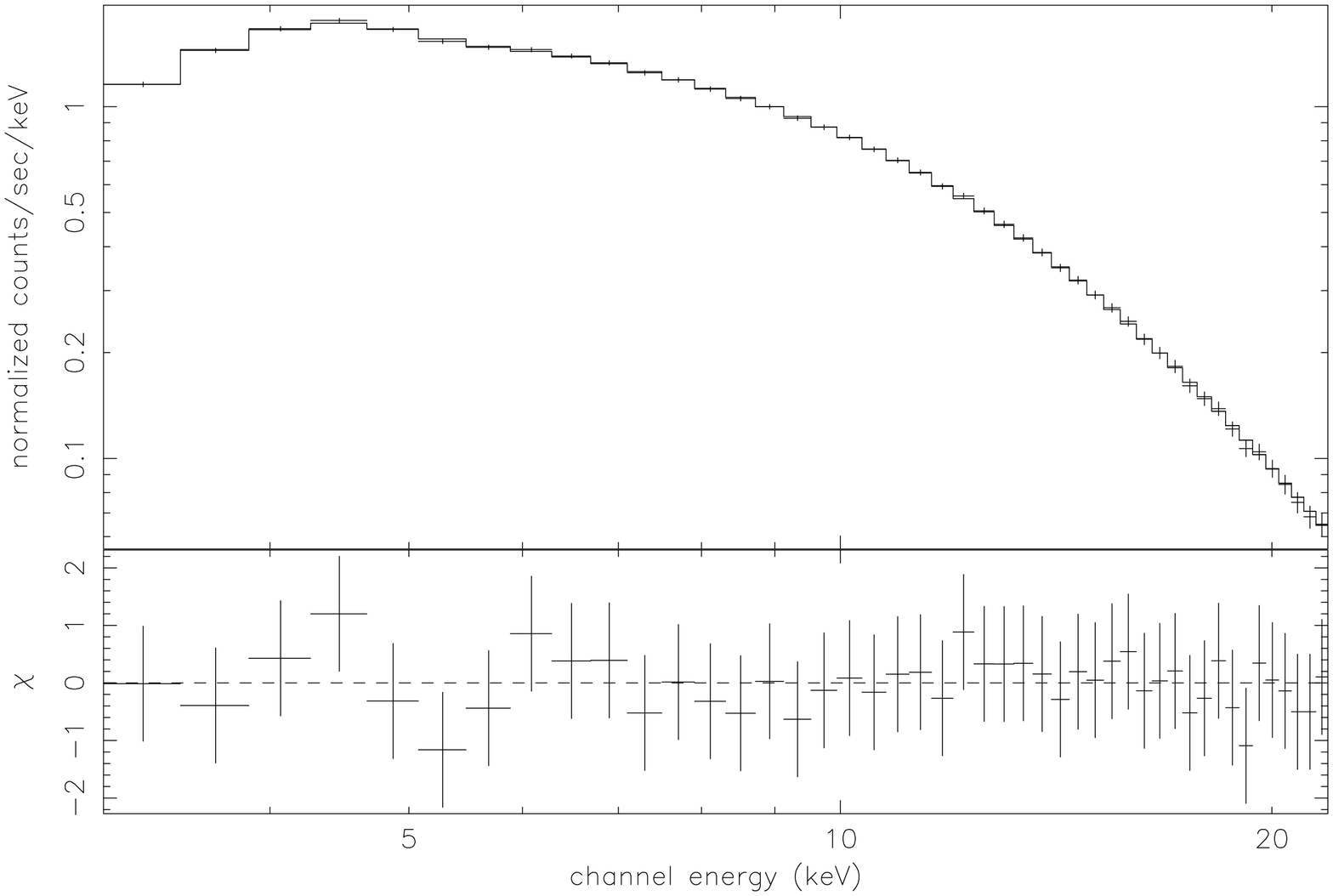}
  \caption{3 -- 22 keV PCA spectrum from all observations combined.}\label{fig2}
\end{figure}

\begin{figure}[h!]
  \includegraphics[height=0.45\textheight]{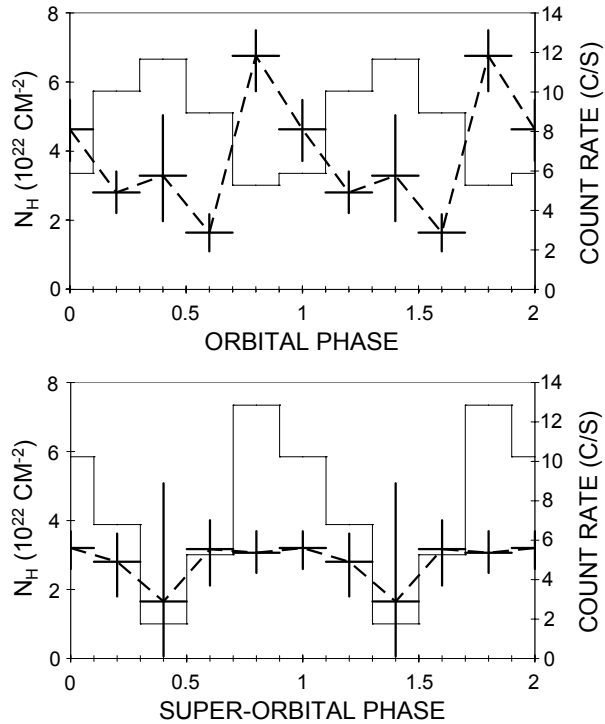}
  \caption{The neutral hydrogen column density from the phase resolved spectral analysis plotted over the orbital (top)
  and super-orbital (bottom) periods. The histograms represent the
  \textit{RXTE} PCA 2 -- 9 keV light curve folded over the respective periods. Phase 0 was arbitrarily
  chosen to coincide with the first \emph{RXTE} observation at MJD 53506.}\label{fig3}
\end{figure}

Between 2005 May 15 -- 2005 June 14 and 2005 December 13 -- 2006
January 12 we obtained 220 ks of pointed observation data of 2S
0114+650 with $\textit{RXTE}$. Two separate runs of 11 observations
were made, each $\sim$3 hr in duration. We extracted separate
spectra from the Standard2f mode data for all 22 observations for
each Proportional Counter Unit (PCU) in the Proportional Counter
Array (PCA) -- excluding PCU0 due to the loss of the propane level
-- combining them to give the best S/N and weighting the response
matrices by exposure (K. Jahoda 2006, private communication). A
1$\%$ systematic error was added to the resulting spectrum to
account for uncertainties in the response matrix. Spectral models
were then fitted to the combined spectrum in the 3 -- 22 keV energy
range using XSPEC v12.3. The best fit model was found to be an
absorbed power law (N$_{H}$ = 2.9 $\pm$ 0.6 $\times$ 10$^{22}$ atoms
cm$^{-2}$; Photon Index = 1.0 $\pm$ 0.1) with a high-energy
exponential cut-off (E$_{CUT}$ = 5.8 $\pm$ 0.5 keV; E$_{FOLD}$ = 14
$\pm$ 2 keV), yielding a model flux of 1.7103 $\times$ 10$^{-10}$
ergs cm$^{-2}$ s$^{-1}$ (Figure~\ref{fig2}). Inclusion of an Fe
K$\alpha$ Gaussian emission line did not improve the fit
significantly. No evidence was found of the $\sim$22 keV cyclotron
absorption feature tentatively identified in \emph{INTEGRAL} data by
\citet{Bonning:2005}. Using the same method we extracted five phase
resolved spectra separated by 0.2 phase bins over the 11.6 d orbital
period. The same absorbed cut-off powerlaw model was fitted to each
of the spectra and the neutral hydrogen column density was plotted
against the orbital period (Figure~\ref{fig3}). The column density
appears to be roughly anti-correlated with the orbital period. The
same phase resolved spectral analysis was repeated over the 30.7 d
super-orbital period. No variability in the column density was found
over the super-orbital period (Figure~\ref{fig3}).

\clearpage

\section{Discussion}

\citet{Migliari:2006} have demonstrated a correlation in neutron
star X-ray binaries between the X-ray (2-10 keV) and radio (8.5 GHz)
luminosity. If their value for the radio luminosity is extrapolated
to 1.38 GHz assuming either a synchrotron power law or a flat
spectrum, their predicted flux for 2S 0114+650 is more than a
magnitude higher than our observed upper limit with the GMRT. Jet
formation, which is linked to radio emission,  may well be
suppressed in 2S 0114+650 by the high neutron star magnetic field.
Furthermore, an accretion disc may not exist in this source. The
observed variability of the neutral hydrogen column density over the
orbital period indicates that variable absorption is the mechanism
behind the orbital modulation, as expected. A lack of variability in
the column density over the super-orbital period indicates that this
modulation is due to a different mechanism, precluding variable
absorption by a warped, precessing accretion disc as the cause of
the super-orbital modulation.


\begin{theacknowledgments}
SAF acknowledges travel support provided by UNSW@ADFA and the
Astronomical Society of Australia to attend the Cefal\`{u} meeting.
We thank the $\textit{RXTE}$ GOF for acquiring the X-ray data and
advice on the data analysis. We thank the GMRT team for the
acquisition of the radio data, M. Pandey for assistance with the
radio data acquisition and data analysis, and P. Lah for assistance
with the radio analysis.
\end{theacknowledgments}



\bibliographystyle{aipproc}   

\bibliography{Farrell_1}

\IfFileExists{\jobname}{}
 {\typeout{}
  \typeout{******************************************}
  \typeout{** Please run "bibtex \jobname" to optain}
  \typeout{** the bibliography and then re-run LaTeX}
  \typeout{** twice to fix the references!}
  \typeout{******************************************}
  \typeout{}
 }

\end{document}